\documentclass[12pt]{article}
\usepackage{graphicx,bm,colonequals,amsmath,amssymb,url,xcolor,bbm}
\usepackage{graphicx,psfrag,epsf}
\usepackage{enumerate}
\usepackage{natbib}
\usepackage{mathtools}
\usepackage[font={footnotesize}]{caption,subcaption}
\mathtoolsset{showonlyrefs=true}
\usepackage[ruled,vlined,noend]{algorithm2e}
\SetKwInput{KwInput}{Input}
\usepackage{algorithmic}

\newcommand{\blind}{0}

\begin{document}

\def\spacingset#1{\renewcommand{\baselinestretch}%
{#1}\small\normalsize} \spacingset{1}

\newcommand{\bh}{\mathbf{h}}
\newcommand{\bu}{\mathbf{u}}
\newcommand{\bl}{\mathbf{l}}
\newcommand{\ba}{\mathbf{a}}
\newcommand{\bv}{\mathbf{v}}
\newcommand{\bc}{\mathbf{c}}
\newcommand{\bb}{\mathbf{b}}
\newcommand{\be}{\mathbf{e}}
\renewcommand{\bf}{\mathbf{f}}
\newcommand{\bs}{\mathbf{s}}
\newcommand{\bp}{\mathbf{p}}
\newcommand{\bx}{\mathbf{x}}
\newcommand{\by}{\mathbf{y}}
\newcommand{\bq}{\mathbf{q}}
\newcommand{\bg}{\mathbf{g}}
\newcommand{\bk}{\mathbf{k}}
\newcommand{\br}{\mathbf{r}}
\newcommand{\bw}{\mathbf{w}}
\newcommand{\bS}{\mathbf{S}}
\newcommand{\bT}{\mathbf{T}}
\newcommand{\bz}{\mathbf{z}}
\newcommand{\bZ}{\mathbf{Z}}
\newcommand{\bO}{\mathbf{O}}
\newcommand{\bP}{\mathbf{P}}
\newcommand{\bA}{\mathbf{A}}
\newcommand{\bY}{\mathbf{Y}}
\newcommand{\bJ}{\mathbf{J}}
\newcommand{\bW}{\mathbf{W}}
\newcommand{\bF}{\mathbf{F}}
\newcommand{\bG}{\mathbf{G}}
\newcommand{\bE}{\mathbf{E}}
\newcommand{\bL}{\mathbf{L}}
\newcommand{\bN}{\mathbf{N}}
\newcommand{\bI}{\mathbf{I}}
\newcommand{\bD}{\mathbf{D}}
\newcommand{\bH}{\mathbf{H}}
\newcommand{\bU}{\mathbf{U}}
\newcommand{\bV}{\mathbf{V}}
\newcommand{\bK}{\mathbf{K}}
\newcommand{\bX}{\mathbf{X}}
\newcommand{\bQ}{\mathbf{Q}}
\newcommand{\bB}{\mathbf{B}}
\newcommand{\bC}{\mathbf{C}}
\newcommand{\bM}{\mathbf{M}}
\newcommand{\bR}{\mathbf{R}}

\newcommand{\bfzero}{\mathbf{0}}
\newcommand{\bfalpha}{\bm{\alpha}}
\newcommand{\bfgamma}{\bm{\gamma}}
\newcommand{\bfmu}{\bm{\mu}}
\newcommand{\bfxi}{\bm{\xi}}
\newcommand{\bftheta}{\bm{\theta}}
\newcommand{\bfeta}{\bm{\eta}}
\newcommand{\bfnu}{\bm{\nu}}
\newcommand{\bfdelta}{\bm{\delta}}
\newcommand{\bfkappa}{\bm{\kappa}}
\newcommand{\bfbeta}{\bm{\beta}}
\newcommand{\bfepsilon}{\bm{\epsilon}}
\newcommand{\bftau}{\bm{\tau}}
\newcommand{\bfomega}{\bm{\omega}}
\newcommand{\bfpi}{\bm{\pi}}
\newcommand{\bfpsi}{\bm{\psi}}
\newcommand{\bfrho}{\bm{\rho}}
\newcommand{\bfSigma}{\bm{\Sigma}}
\newcommand{\bfGamma}{\bm{\Gamma}}
\newcommand{\bfLambda}{\bm{\Lambda}}
\newcommand{\bfPsi}{\bm{\Psi}}
\newcommand{\bfOmega}{\bm{\Omega}}

\newcommand{\order}{\mathcal{O}}


\if0\blind
{
  \title{\textbf{Scalable Sampling of Truncated Multivariate Normals Using Sequential Nearest-Neighbor Approximation}}
  \author{Jian Cao
  \\
    Department of Mathematics, University of Houston\\
    and \\
    Matthias Katzfuss \\
    Department of Statistics, University of Wisconsin--Madison}
  \maketitle
} \fi

\if1\blind
{
  \bigskip
  \bigskip
  \bigskip
  \begin{center}
    {\LARGE\bf Title}
\end{center}
  \medskip
} \fi

\bigskip
\begin{abstract}
We propose a linear-complexity method for sampling from truncated multivariate normal (TMVN) distributions with high fidelity by applying nearest-neighbor approximations to a product-of-conditionals decomposition of the TMVN density. To make the sequential sampling based on the decomposition feasible, we introduce a novel method that avoids the intractable high-dimensional TMVN distribution by sampling sequentially from $m$-dimensional TMVN distributions, where $m$ is a tuning parameter controlling the fidelity. This allows us to overcome the existing methods' crucial problem of rapidly decreasing acceptance rates for increasing dimension. Throughout our experiments with up to tens of thousands of dimensions, we can produce high-fidelity samples with $m$ in the dozens, achieving superior scalability compared to existing state-of-the-art methods. We study a tetrachloroethylene concentration dataset that has $3{,}971$ observed responses and $20{,}730$ undetected responses, together modeled as a partially censored Gaussian process, where our method enables posterior inference for the censored responses through sampling a $20{,}730$-dimensional TMVN distribution.
\end{abstract}

\noindent%

\spacingset{1.45}
\section{Introduction}
\label{sec:intro}

Truncated multivariate normal (TMVN) distributions are used in a variety of applications, for example, in the modelling of binary responses \citep{cao2022scalable}, extreme events \citep{Nascimento2022}, and skew-normal variables \citep{zhang2010spatial, cao2021exploiting}, and Gaussian process (GP) regression with inequality constraints \citep{da2012gaussian}, for which posterior inference largely depends on the capability of sampling from a TMVN distribution. 

The gold standard for sampling from TMVN distributions is the accept-reject method based on the exponentially tilted normal proposal density introduced in \cite{botev2017normal}. This proposal density is constructed based on the separation-of-variable (SOV) transformation of the multivariate normal (MVN) density function introduced in \cite{genz1992numerical} and achieves significant improvement on the acceptance rate for drawing TMVN samples, especially when the TMVN distribution is truncated away from the mean. Despite significantly higher acceptance rates than the SOV method, \cite{botev2017normal} is not the panacea for increasing dimensions of TMVN distributions, denoted by $n$. Empirically, when $n > 1{,}000$, the acceptance rate of \cite{botev2017normal} becomes too small to be computationally feasible. Based on \cite{botev2017normal}, \cite{cao2023linear} adapt the Vecchia Gaussian-process (GP) approximation \citep{Vecchia1988} to reduce the complexity per Monte Carlo (MC) sample from the proposal density from $\order(n^2)$ to $\order(n)$, but the reduced complexity is not sufficient to compensate for the rapidly decreasing acceptance rate with increasing dimension $n$. \cite{sahoo2024computationally} tackles the issue of low acceptance rate by assuming independence between the censored responses conditional on a latent GP realization over a fine mesh covering the spatial domain of the censored responses and uses Markov chain Monte Carlo (MCMC) to generate samples of censored responses. However, this assumption may cause significant deviation between the generated samples and the target TMVN distribution. 

We believe that the crux of sampling high-dimensional TMVN distributions is overcoming the rapidly decreasing acceptance rate as $n$ increases while maintaining a high fidelity between the generated samples and the target TMVN distribution. In this paper, we extend the idea of applying the Vecchia approximation to the MVN density as proposed in \cite{cao2023linear} to applying a sequential nearest-neighbor approximation to TMVN densities and propose a novel method for sampling marginal TMVN distributions. 
Our methods are most appropriate for covariance matrices obtained by evaluating a kernel at pairs of inputs or locations, so that neighbors are naturally defined in the input space; however, our method can also be applied to generic covariance matrices, for example via a distance defined by correlation \citep{kang2023correlation}. 
Our idea also shares similarity with \citet{Nascimento2022} in terms of truncating the interval-type conditions (e.g., $\bx \in [\ba, \bb]$), but the latter applies the truncation to a totally different decomposition aiming to estimate MVN probabilities. Our proposed method breaks down the sampling of a $n$-dimensional TMVN distribution into a series of $m$-dimensional TMVN samplings, maintaining a constant acceptance rate for increasing $n$, controlled by the tuning parameter $m$. The generated samples have a high fidelity to the target TMVN distribution. Furthermore, our proposed method has linear complexity with respect to $n$, making it much more scalable than other state-of-the-art methods. 

In this paper, we showcase an important application of sampling TMVN distributions: posterior inference for partially censored GPs. GPs are extensively used for modeling data collected by sensors distributed over a spatial domain \citep{chen2021space, salvana20223d, adak2023phenomic}. Partially censored GPs arise, for example, in the case of unmeasured responses due to detection limits. It is important to incorporate the information of the censored responses into posterior inference, for which state-of-the-art methods hinge on generating high-quality samples from the target TMVN distribution. 

The paper is organized as follows: Section~\ref{sec:meth} describes our proposed method; Section~\ref{sec:numer_cmp} demonstrates the high fidelity and scalability of the proposed method; Sections~\ref{sec:sim_study} and \ref{sec:app_study} use a simulated high-dimensional partially censored GP realization and a real tetrachloroethylene concentration dataset, respectively, to compare the proposed method with the state of the art in terms of posterior inference. Section~\ref{sec:conc} concludes the paper.

\section{Methods}
\label{sec:meth}

Suppose $\by$ is a random vector from a $n$-dimensional TMVN distribution $\mathcal{TN}(\ba, \bb; \bfSigma)$, where $\bfSigma$ is a covariance matrix and $\ba$ and $\bb$ are the lower and upper bounds, respectively, defining a hyper-rectangle in $\mathbb{R}^{n}$ into which the MVN distribution $N(\bfzero, \bfSigma)$ is truncated. Our proposed method is developed in the context of kernel-based covariances, where each response $y_i$ is associated with an input or location $\bs_{i} \in \mathbb{R}^{d}$ and the covariance matrix $\bfSigma$ is computed with a covariance kernel to facilitate the nearest-neighbor construction to be described below; however, our method can be extended to generic covariance matrices, for example via a distance defined by correlation \citep{kang2023correlation}. Specifically, the covariance matrix $\bfSigma$ is computed based on the locations $\{\bs_{i} \in \mathbb{R}^{d}\}_{i = 1}^{n}$ and a covariance kernel $\mathcal{K}$, such that $\bfSigma_{i, j} = \mathcal{K}(\bs_{i}, \bs_{j})$. 

Mathematically, $\by$ has the same distribution as $\bz \mid \ba \le \bz \le \bb$, where $\bz \sim N(\bfzero, \bfSigma)$, and vector inequalities are considered to apply elementwise for all entries. Using $f(\cdot)$ to denote a generic density function, we have
\begin{align}
    \label{equ:cond_decomp}
    f(\by) = f(z_{1} | \ba \le \bz \le \bb) f(z_{2} | z_{1}, \ba \le \bz \le \bb) \cdots f(z_{n} | \bz_{1:n-1}, \ba \le \bz \le \bb) .
\end{align}
Unlike MVN distributions, the conditional components on the right-hand-size of \eqref{equ:cond_decomp} do not have an analytical form, which makes sequential sampling based on \eqref{equ:cond_decomp} infeasible.

The first part of our proposed method is to use nearest neighbors to reduce the number of conditioning inequalities in each component of \eqref{equ:cond_decomp} to $m$. Specifically, define $c(i)$ as the collection of indices of the $m$ nearest neighbors of $\bs_i$, which is the location corresponding to $y_{i}$. Assume that the first entry of $c(i)$ is $i$. Denote by $c^{p}(i) = \{1, \ldots, i - 1\} \bigcap c(i)$ and $c^{l}(i) = c(i) \backslash c^{p}(i)$ the previous and later-ordered indices, respectively.
We propose a nearest-neighbor approximation of the form
\begin{align}
f(z_{i} | \bz_{1:i-1}, \ba \le \bz \le \bb) \approx f(z_{i} | \bz_{c^{p}(i)}, \ba_{c(i)} \le \bz_{c(i)} \le \bb_{c(i)}) = f(z_{i} | \bz_{c^{p}(i)}, \ba_{c^{l}(i)} \le \bz_{c^{l}(i)} \le \bb_{c^{l}(i)}). \label{equ:component_approx}
\end{align}
Overall, this results in the approximation
\begin{align}
    f(\by) \approx{} &f(z_{1} | \ba_{c^{l}(1)} \le \bz_{c^{l}(1)} \le \bb_{c^{l}(1)}) f(z_{2} | z_{c^{p}(2)}, \ba_{c^{l}(2)} \le \bz_{c^{l}(2)} \le \bb_{c^{l}(2)}) \cdots \\
    \label{equ:cond_decomp_approx}
    &f(z_{n} | \bz_{c^{p}(n)}, \ba_{c^{l}(n)} \le \bz_{c^{l}(n)} \le \bb_{c^{l}(n)}).
\end{align}
The intuition for this nearest-neighbor approximation is to exploit the spatial Markovian property or the `screening effect' \citep{Stein2011} that assumes the impact of far-away responses on computing the conditioning distribution is minimal after conditioning on nearby responses. Since \eqref{equ:cond_decomp_approx} indicates a sequential sampling scheme, we name our method sequential-nearest-neighbor (SNN) sampling. 

However, sampling from the right-hand side of \eqref{equ:component_approx} is also challenging. Thus, the second component of our proposed method is to view $f(z_{i} | \bz_{c^{p}(i)}, \ba_{c^{l}(i)} \le \bz_{c^{l}(i)} \le \bb_{c^{l}(i)})$ as a marginal distribution of a $|c^{l}(i)|$-dimensional TMVN distribution, $f(\bz_{c^{l}(i)} | \bz_{c^{p}(i)}, \ba_{c^{l}(i)} \le \bz_{c^{l}(i)} \le \bb_{c^{l}(i)})$, whose mean and covariance matrix are essentially the conditional mean and covariance of $\bz_{c^{l}(i)} | \bz_{c^{p}(i)}$. In other words, to generate a sample from $f(z_{i} | \bz_{c^{p}(i)}, \ba_{c^{l}(i)} \le \bz_{c^{l}(i)} \le \bb_{c^{l}(i)})$, one can sample
\[
\tilde{\bz}_{c^{l}(i)} \sim f(\bz_{c^{l}(i)} | \bz_{c^{p}(i)}, \ba_{c^{l}(i)} \le \bz_{c^{l}(i)} \le \bb_{c^{l}(i)}),
\]
and keep only the first entry of $\tilde{\bz}_{c^{l}(i)}$. It is worth mentioning that $f(\bz_{c^{l}(i)} | \bz_{c^{p}(i)}, \ba_{c^{l}(i)} \le \bz_{c^{l}(i)} \le \bb_{c^{l}(i)})$ may be very different from $f(\bz_{c^{l}(i)} | \bz_{1 : i - 1}, \ba \le \bz \le \bb)$, but the marginal distribution of $z_{i}$ indicated by the former is highly aligned with that in the latter, which will be shown in later sections. 
\begin{algorithm}[ht!]
\caption{The SNN Method for drawing TMVN samples}
\label{alg:SNN}
\KwInput{Parameters $\ba, \bb, \bfSigma$; locations $\{\bs_{i}\}_{i = 1}^{n}$; tuning parameter $m$; number of samples $N$}
\KwResult{$N$ (approximate) samples from the target TMVN distribution $\mathcal{TN}(\ba, \bb; \bfSigma)$}
\begin{algorithmic}[1]
\STATE (Optional) Reorder $\ba, \bb, \bfSigma$ according to a maximin ordering of $\{\bs_{i}\}_{i = 1}^{n}$
\FOR{$i = 1, 2, \ldots, n$}
    \STATE Find the $m$ nearest neighbors of $\bs_{i}$, denoted by $c(i)$
    \STATE Define $c^{p}(i) = \{1, \ldots, i - 1\} \bigcap c(i)$ and $c^{l}(i) = c(i) \backslash c^{p}(i)$
    \STATE Compute $\bV_{i} = \bfSigma_{c^{l}(i), c^{p}(i)} \bfSigma_{c^{p}(i), c^{p}(i)}^{-1}$, $\tilde{\bfSigma}_{i} = \bfSigma_{c^{l}(i), c^{l}(i)} - \bV_{i} \bfSigma_{c^{p}(i), c^{l}(i)}$
\ENDFOR
\FOR{$k = 1, 2, \ldots, N$}
\FOR{$i = 1, 2, \ldots, n$}
  \STATE Compute $\tilde{\bfmu}_{i}^{(k)} = \bV_{i} \by^{(k)}_{c^{p}(i)}$
  \STATE Draw a sample $\tilde{\bz}_{i}^{(k)}$ from $\mathcal{TN}(\ba_{c^{l}(i)} - \tilde{\bfmu}_{i}^{(k)}, \bb_{c^{l}(i)} - \tilde{\bfmu}_{i}^{(k)}; \tilde{\bfSigma}_{i})$ \label{stp:draw_samp}
  \STATE Set $y_{i}^{(k)}$ to the first entry of $\tilde{\bz}_{i}^{(k)} + \tilde{\bfmu}_{i}^{(k)}$
\ENDFOR
\ENDFOR
\RETURN $\{\by^{k}\}_{k = 1}^{N}$
\end{algorithmic}
\end{algorithm}
Our proposed SNN method is outlined in Algorithm~\ref{alg:SNN}. We sample $\tilde{\bz}_{c^{l}(i)}$ from the $|c^{l}(i)|$-dimensional TMVN distribution on Line~\ref{stp:draw_samp} using minimax exponential tilting \citep{botev2017normal}, the gold standard for sampling low-dimensional TMVN distributions with, say, $n<100$, which aligns with the commonly used magnitude of the tuning parameter $m$. Our proposed SNN method has a complexity of $\order(nm^{3})$ and can be trivially parallelized across the generation of $N$ TMVN samples.

The proposed method SNN solves the challenge of other methods' extremely low acceptance rate when $n$ is larger than a few hundred, by restricting the TMVN sampling dimension to no greater than $m$ and maintains a high fidelity in the approximation of each conditional density in \eqref{equ:cond_decomp_approx} based on the screening effect. Although the screening effect for TMVN distributions has not been as comprehensively tested as that for GPs, \cite{cao2023linear} provided numerical support for the screening effect by showing that the generated samples for the locations of interest had minimal difference whether considering all locations over which the TMVN distribution is defined or a smaller subset enveloping the locations of interest. Furthermore, for the simulation and real-data studies in this paper, the SNN approximation demonstrates high fidelity. 

SNN is a sequential sampling method, for which different orderings of the responses $\by$ may result in different distributional errors when approximating \eqref{equ:cond_decomp} with \eqref{equ:cond_decomp_approx}. \cite{Schafer2020} showed that the maximum-minimum distance (maximin) ordering achieved $\epsilon$-accuracy when applying the Vecchia approximation \citep{Vecchia1988} to one specific family of GPs (i.e.,  Green’s functions of elliptic boundary-value). Furthermore, maximin ordering has been shown to reduce the errors of the Vecchia approximations of various other GPs \citep{Katzfuss2017a, katzfuss2020vecchia, Guinness2018}. Similar to \eqref{equ:cond_decomp_approx}, the Vecchia approximation also uses nearest responses to reduce the conditioning sets based on the screening effect. In this paper, the SNN method, by default, uses the coordinate-oriented ordering (e.g., $x$ and $y$-coordinates being the primary and secondary keys, respectively) for locations on a grid and random orderings otherwise. We use the acronym MSNN for applying the maximin ordering to $\by$ and $\{\bs\}_{i = 1}^{n}$ prior to the SNN method (Line 1 in Algorithm~\ref{alg:SNN}) and find that it heuristically improves SNN in high dimensions (e.g., $n > 1{,}000$).

\section{Numerical Exploration of Scalability and Fidelity of SNN}
\label{sec:numer_cmp}

In this section, we first consider some low-dimensional TMVN distributions with $n<500$ to compare the deviation of the samples generated by the Vecchia minimax exponential tilting \citep[VMET;][]{cao2023linear}, Bayesian SPDE for censored GP \citep[CSB;][]{sahoo2024computationally}, and our proposed SNN methods to those generated by the gold standard of the minimax exponential tilting \citep[MET;][]{botev2017normal} method. Next, we examine the scalability of the methods mentioned above; by gradually increasing the dimensions of the TMVN distributions, we demonstrate the exponentially decreasing acceptance ratios of MET and VMET, which makes them computationally infeasible for moderately large $n$. The Mat\'ern covariance kernel is a popular covariance kernel, unifying several well-known kernels as special cases, including exponential, Whittle, and squared exponential kernels, and so the Mat\'ern family is used for numerical examples in this paper. 

We considered partially censored GPs, which can be viewed as a limiting case of TMVN distributions. Specifically, a partially censored GP can be viewed as $\mathcal{TN}(\ba, \bb; \bfSigma)$ with $\ba_{\mathcal{O}} = \bb_{\mathcal{O}} = \bz_{\mathcal{O}}$, where $\mathcal{O}$ is the index set for observed responses (which are assumed to be ordered first).
In the first experiment, we generated responses from a GP $\bz$ over $n = 400$ spatial locations, distributed as a $20 \times 20$ grid in the unit square, under zero mean and an isotropic Mat\'ern-$1.5$ kernel with variance and range equal to $1.0$ and $0.1$, respectively. Responses below $\bb = \mathbf{1}$ were censored, which led to $n^{o} = 70$ observed responses and $n^{c} = 330$ censored responses. We slightly abuse notation, in that when discussing partially censored GPs, $n$ denotes the total number of responses and $n^{c}$ denotes the non-trivial dimensions of TMVN distributions.
Algorithm~\ref{alg:SNN} can be directly applied to drawing samples from a partially censored GP.  For MET, we computed the conditional MVN distribution for the censored responses and used the \texttt{TruncatedNormal} \citep{botev2017normal} R package to generate $N = 50$ samples of the $330$-dimensional TMVN distribution. For VMET and CSB, the partially censored GP responses could be directly used as inputs to the \texttt{VeccTMVN} \citep{cao2023linear} and the \texttt{CensSpBayes} \citep{sahoo2024computationally} R packages, respectively, to simulate the underlying GP responses at the $330$ locations with censored responses.
\begin{table}[ht!]
    \centering
    \begin{tabular}{c|c|c|c|c|c}
        \hline
         & CSB & SNN & MSNN & MET & VMET\\
        \hline
        RMSE & 0.54 & 0.43 & 0.46 & 0.43 & 0.44\\
        \hline
        CRPS & 0.29 & 0.23 & 0.24 & 0.23 & 0.23\\
        \hline
    \end{tabular}
    \caption{RMSE and CRPS for predicting $330$ censored GP responses based on $70$ observed GP responses, assuming that the covariance structure and the censoring levels are known}
    \label{tbl:lowdim_exp}
\end{table} 
CSB method used a burn-in size of $2{,}000$ and a thinning factor of $5$, which were the default values in the \texttt{CensSpBayes} R package. Both VMET and (M)SNN set the tuning parameter $m$ to $30$, representing the sizes of conditioning sets and nearest neighbors, respectively. 

\begin{figure}[ht!]
\centering
\begin{subfigure}{.32\textwidth}
\centering
\includegraphics[width =.99\linewidth]{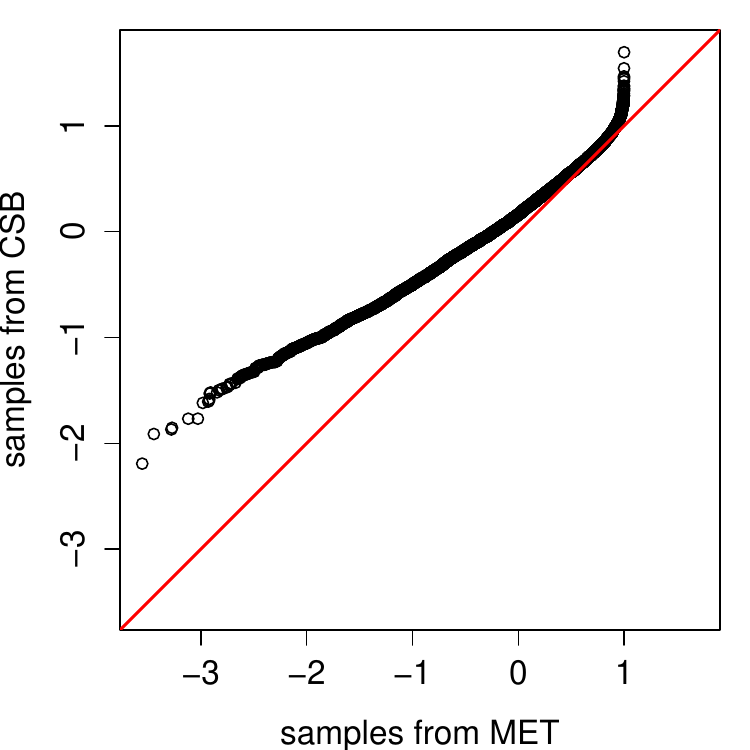}
\end{subfigure}%
\hfill
\begin{subfigure}{.32\textwidth}
\centering
\includegraphics[width =.99\linewidth]{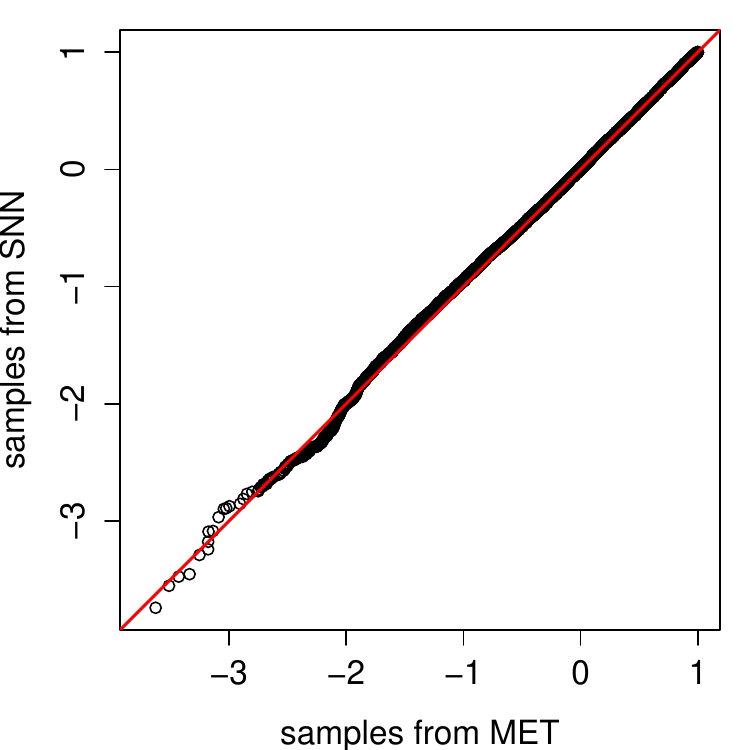}
\end{subfigure}%
\hfill
\begin{subfigure}{.32\textwidth}
\centering
\includegraphics[width =.99\linewidth]{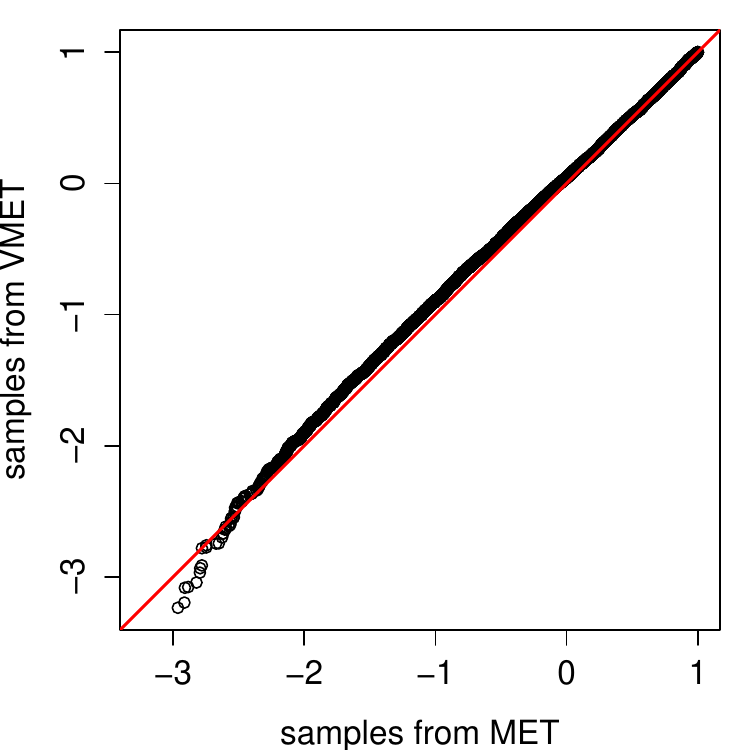}
\end{subfigure}%
\caption{QQ-plots against TMVN samples generated with the gold-standard MET method \citep{botev2017normal}. $50$ samples of the same $330$-dimensional TMVN target distributions were used for comparison. From left to right, samples were generated using CSB \citep{sahoo2024computationally}, SNN, and VMET \citep{cao2023linear}.}
\label{fig:lowdim_scene1}
\end{figure}

Table~\ref{tbl:lowdim_exp} summarizes the root-mean-square error (RMSE) and the continuous ranked probability score (CRPS) for predicting the underlying GP responses at locations with censored responses using CSB, SNN, MSNN, MET, and VEMT.
For this low-dimensional experiment, all methods except for CSB had comparable accuracy, while CSB had slightly higher RMSE and CRPS than other methods, highlighting that the approximated distribution underlying SNN was more accurate than that underlying CSB, and in fact, close to the truth. We think that the difference between SNN and MSNN from Table~\ref{tbl:lowdim_exp} was not significant enough to draw conclusions on whether maximin ordering benefits SNN or not. Figure~\ref{fig:lowdim_scene1} compared the distributions of the TMVN samples generated by CSB, SNN, and VMET, benchmarked against those generated by MET.
The samples generated by CSB had a significant disagreement with those generated by MET, whereas SNN and VMET were both highly aligned with the benchmark MET. The qq-plot between MSNN and MET was essentially the same as that between SNN and MET, and is hence not shown.

The second experiment considered a series of TMVN distributions of increasing dimensions with spatial covariance matrices defined over 2D grids with a spacing of $0.02$. The TMVN distributions had zero-mean structure and their covariance kernel remained a Mat\'ern-$1.5$ kernel but the range parameter was reduced to $0.03$ to avoid numerical singularity. The lower and upper bounds $\ba$ and $\bb$ were defined as $-\mathbb{\infty}$ and $\mathbf{0}$, respectively.
\begin{figure}[ht!]
\centering
\hspace*{\fill}
\begin{subfigure}{.45\textwidth}
\centering
\includegraphics[width =.99\linewidth]{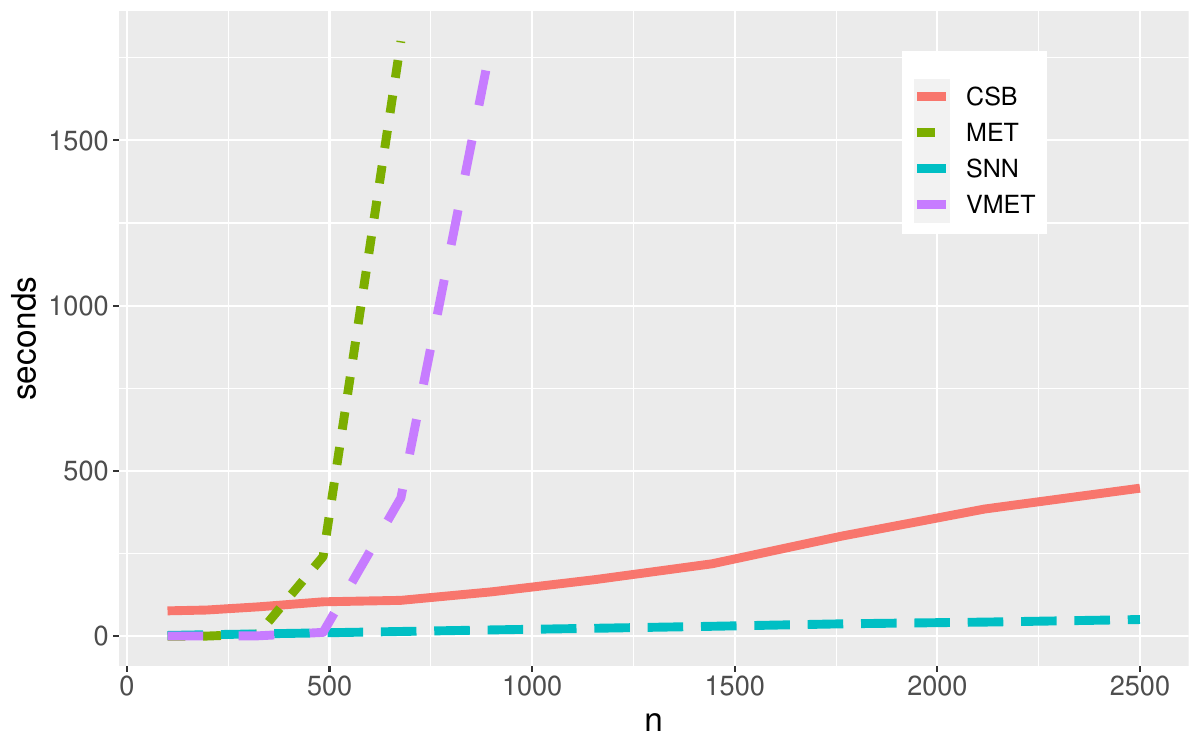}
\end{subfigure}%
\hfill
\begin{subfigure}{.45\textwidth}
\centering
\includegraphics[width =.99\linewidth]{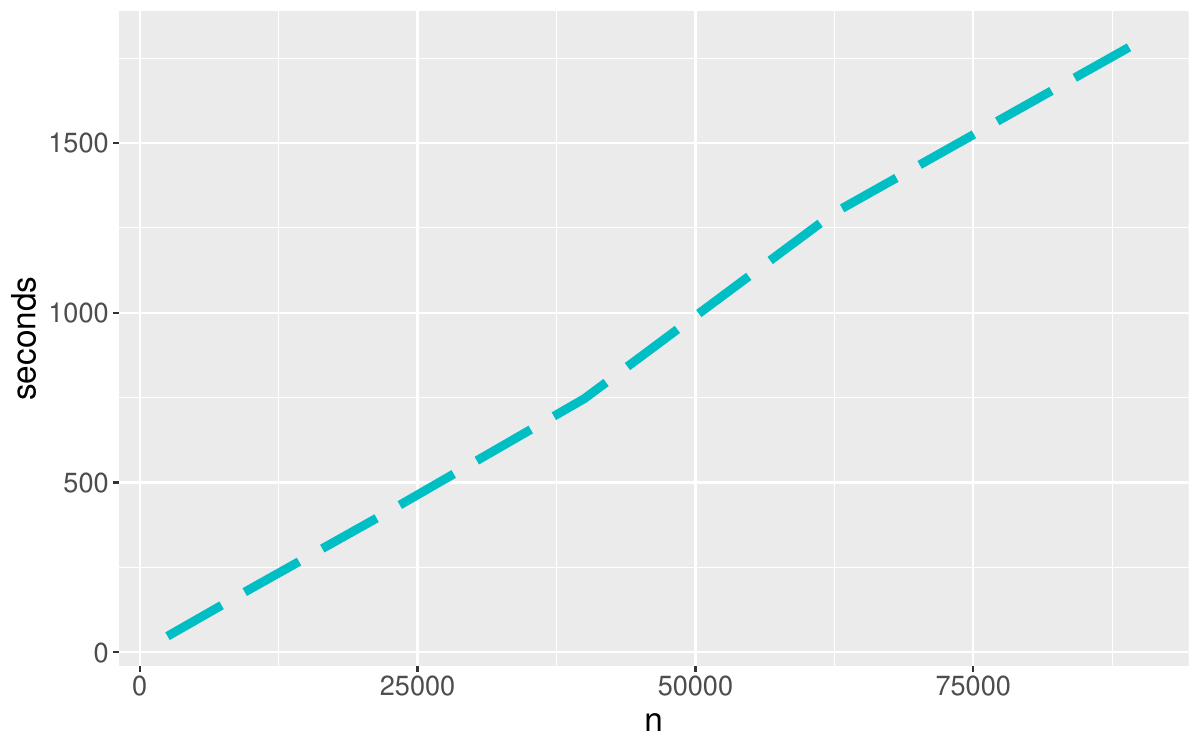}
\end{subfigure}%
\hspace*{\fill}
\caption{Wall time for generating $10$ samples from a $n$-dimensional TMVN distribution using all four methods (left) and SNN only (right).}
\label{fig:time_cmp}
\end{figure}
The wall times for generating $10$ samples from the $n$-dimensional TMVN distribution using CSB, SNN, VMET, and MET are shown in Figure~\ref{fig:time_cmp}, where the cut-off time was $30$ minutes. All methods were measured on the same scientific workstation using a single core, and the tuning parameters for CSB, SNN, and VMET remained the same as in the first experiment. Both VMET and MET suffered from the curse of dimensionality, with acceptance rates decreasing exponentially, becoming computationally infeasible for moderately large $n$ (e.g., $900$). (To tackle the decreasing acceptance rate, \cite{cao2023linear} proposed a partition technique that only draws TMVN samples over a subset of the domain, enveloping the area of interest.) CSB was overall computationally feasible although its growth trend indicated a super-linear complexity, whereas the computing time for our proposed SNN increased exactly linearly with $n$. We also probed the capability of SNN up to $n \approx 10^5$, significantly extending the computation limits in the literature, to showcase the linear scalability of SNN. It is worth mentioning that CSB requires building a mesh enveloping the spatial locations $\{\bs_{i}\}_{i}^{n}$ in $\mathbb{R}^{d}$, which may become infeasible for $d \ge 3$ \citep{lindgren2023fmesher} whereas SNN depends on $d$ only through the conditioning sets $\{c(i)\}_{i = 1}^{n}$, which has been shown scalable to $d$ on the order of hundreds to thousands \citep{cao2022scalable, cao2023variational}. Maximin ordering has a quasilinear complexity with respect to $n$ and amounts to a minor cost compared with SNN that uses a default coordinate-oriented ordering. Hence, the growth curve of MSNN is not shown in Figure~\ref{fig:time_cmp}.

\section{High-Dimensional Simulated Partially Censored GP}
\label{sec:sim_study}

We also considered a more realistic simulation scenario where the covariance structure is unknown and the number of censored responses are too large for VMET and MET to simulate jointly. We compared the posterior inference and the generated responses at censored locations using our proposed SNN and the existing CSB methods. Additionally, we used the `partition' technique introduced in \cite{cao2023linear} to generate samples independently over partitions of the entire region using VMET and MET as comparisons. However, due to the partition technique, the samples produced by MET and VMET have discontinuities at the boundaries between partitions.

The dataset was generated as GP responses over a $100 \times 100$ grid in $[0, 1]^{2}$ with a zero-mean structure and a Mat\'ern-$1.5$ kernel with variance, range, and nugget equal to $(1.0, 0.03, 10^{-4})$. GP responses below $\bb = \mathbf{1}$ were censored and the censoring threshold $\bb$ was assumed known, which led to $n^{c} = 8{,}515$ censored responses. For covariance estimation, we used the same method as \cite{cao2023linear}, optimizing the log-likelihood of a partially censored GP under the Vecchia approximation that is implemented in the \texttt{VeccTMVN} R package \citep{cao2023linear}. Posterior inference using CSB, SNN, and MSNN with estimated parameters is summarized in Table~\ref{tbl:highdim_exp_parmest}.
\begin{table}[ht!]
    \centering
    \begin{tabular}{c|c|c|c}
        \hline
         & CSB & SNN & MSNN \\
        \hline
        RMSE & 0.83 & 0.61 & 0.58 \\
        CRPS & 0.45 & 0.33 & 0.32 \\
        \hline
    \end{tabular}
    \caption{RMSE and CRPS for the posterior inference at locations with censored GP responses, where the covariance structure is estimated.}
    \label{tbl:highdim_exp_parmest}
\end{table}
For CSB, we used a total MCMC iteration of $25{,}000$, with $20{,}000$ burn-in and a thinning factor of $5$, which was the same as those in \cite{sahoo2024computationally}. SNN and MSNN each generated $50$ samples at locations with censored responses, both achieving significantly higher inference accuracy than CSB. Maximin ordering slightly improved posterior inference as indicated by the score differences between SNN and MSNN. On the same scientific workstation, (M)SNN used four cores and a total of 165 seconds while CSB used $8{,}567$ seconds.

\begin{figure}[ht!]
\centering
\begin{subfigure}{.32\textwidth}
\centering
\includegraphics[width =.99\linewidth]{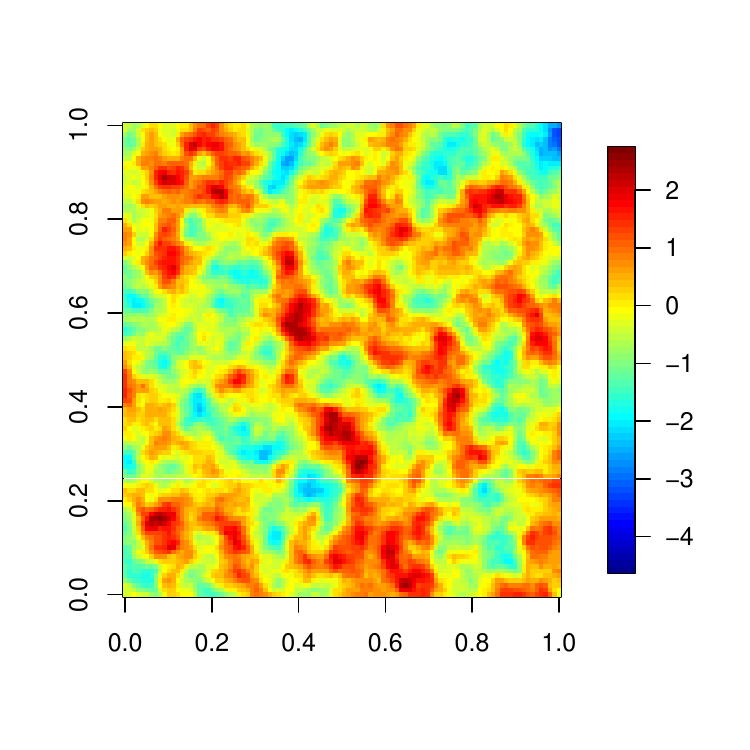}
\vspace{-2\baselineskip}
\caption{Truth}
\end{subfigure}%
\hfill
\begin{subfigure}{.32\textwidth}
\centering
\includegraphics[width =.99\linewidth]{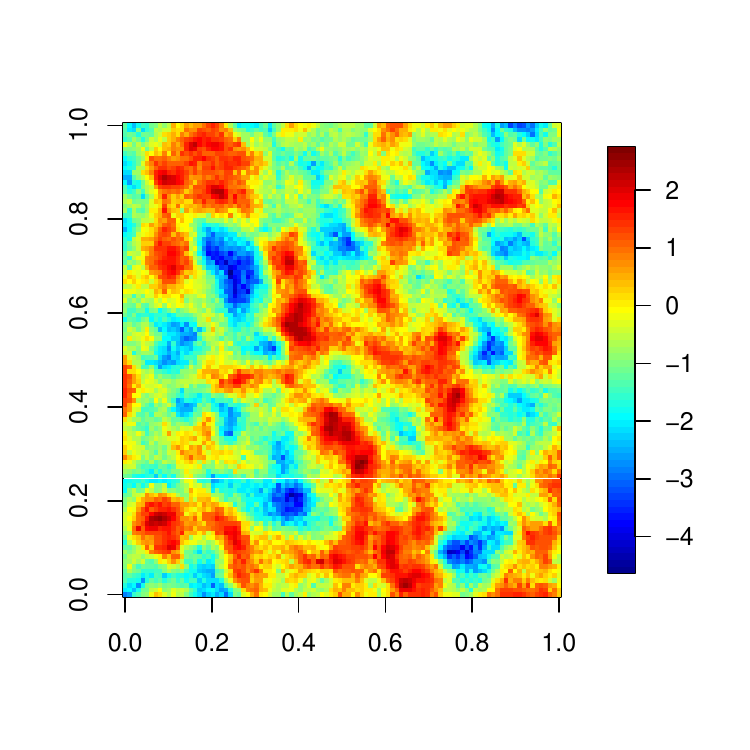}
\vspace{-2\baselineskip}
\caption{CSB}
\end{subfigure}%
\hfill
\begin{subfigure}{.32\textwidth}
\centering
\includegraphics[width =.99\linewidth]{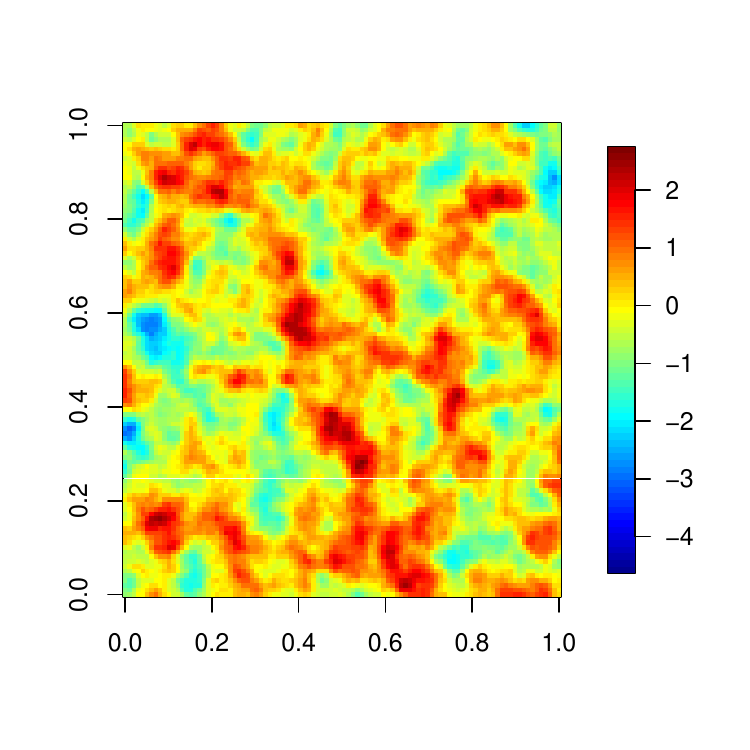}
\vspace{-2\baselineskip}
\caption{SNN}
\end{subfigure}%
\hspace*{\fill} \\

\hfill
\begin{subfigure}{.32\textwidth}
\centering
\includegraphics[width =.99\linewidth]{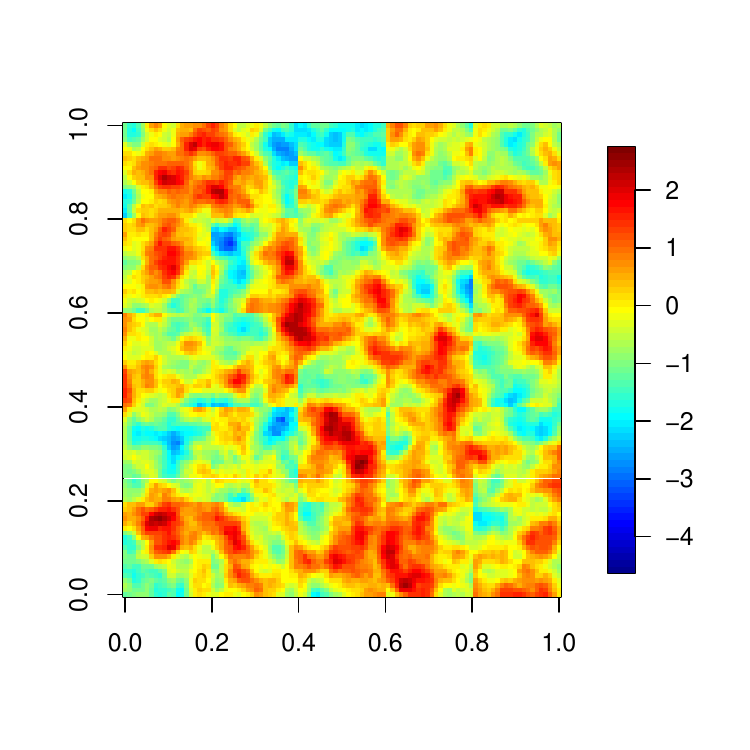}
\vspace{-2\baselineskip}
\subcaption{VMET}
\end{subfigure}
\hfill
\begin{subfigure}{.32\textwidth}
\centering
\includegraphics[width =.99\linewidth]{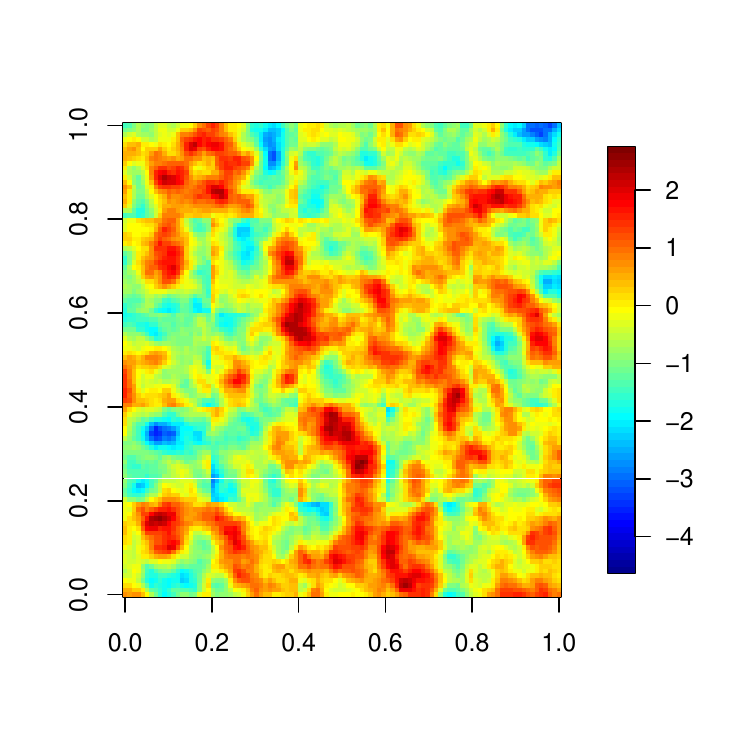}
\vspace{-2\baselineskip}
\subcaption{MET}
\end{subfigure}
\hspace*{\fill} \\

\caption{Heatmaps of the true GP field and the samples generated by CSB \citep{sahoo2024computationally}, our proposed SNN, VMET \citep{cao2023linear}, and MET \citep{botev2017normal}}
\label{fig:sim_study}
\end{figure}

Next, we used VMET and MET, both combined with the partition technique, to predict the censored GP responses. Here, we assumed the true covariance parameters were known to showcase the accuracy gain from knowing the true covariance. The results are summarized in Table~\ref{tbl:highdim_exp_parmknown}.
\begin{table}[ht!]
    \centering
    \begin{tabular}{c|c|c|c|c}
        \hline
          & SNN & MSNN & MET & VMET\\
        \hline
        RMSE & 0.58 & 0.58 & 0.57 & 0.58\\
        CRPS & 0.31 & 0.31 & 0.31 & 0.31\\
        \hline
    \end{tabular}
    \caption{RMSE and CRPS for the posterior inference at locations with censored GP responses, where the true covariance structure is known.}
    \label{tbl:highdim_exp_parmknown}
\end{table}
Since we could not encode the `true' covariance structure into the CSB method, CSB was not included. Here, the partition evenly divided the spatial domain ($[0, 1]^2$) into $5 \times 5$ smaller squares and sampled the TMVN distribution defined over each smaller square independently with an offset of $0.05$. With the partition technique, VMET and MET predicted the underlying GP responses at each censored location roughly as accurately as SNN, because the partition technique is essentially based on the screening effect that is also the underlying intuition of SNN. However, joint posterior inference using VMET and MET is not possible using this partition technique. Figure~\ref{fig:sim_study} shows visualizations of the true underlying GP and four samples generated by CSB, SNN, VMET, and MET, respectively, where VMET and MET had discontinuities at the boundaries between partitions. Such artifacts often make the results unattractive to partitioners.
The samples generated by SNN and MSNN, with either estimated or true covariance parameters, were visually indistinguishable, and so we only displayed the heatmap of a sample generated by SNN with the true covariance. Compared with other methods, CSB had an overall negative bias and a rougher spatial structure.

\section{Application To Partially Censored Groundwater Data}
\label{sec:app_study}

We demonstrate the capability of our proposed SNN for jointly sampling a large number of censored GP responses on a dataset of tetrachloroethylene concentrations from the United States Geological Survey (USGS) over the time period 2000 to 2022, which was also analyzed in \cite{cao2023linear}. Tetrachloroethylene can be used in dry cleaning, manufacturing other chemicals, and cleaning metals, but exposure to tetrachloroethylene is harmful and may cause skin and respiratory problems. The dataset consists of $n = 24{,}701$ spatio-temporal responses, among which $n^{c} = 20{,}730$ are left-censored due to the detection capability of the sensors. The censoring threshold or the limit of detection (LOD), $\bb$, are known and different across spatio-temporal locations. 
Environmental contaminants are usually modeled as a logarithm GP \citep{helsel1990less}. As in \cite{cao2023linear}, we used a partially censored zero-mean GP with a Mat\'ern-1.5 kernel to model the normalized logarithms of the tetrachloroethylene concentration and the LODs. Different from \cite{cao2023linear}, we assigned three separate range parameters to the three physical dimensions of longitude, latitude and time.

We optimized the log-likelihood of a partially censored GP through the \texttt{VeccTMVN} R package \citep{cao2023linear} to estimate the covariance parameters, which produced a variance estimate of $8.76$, range estimates of $(0.09, 0.15, 10^{6})$, corresponding to longitude, latitude, and time, respectively, and a nugget estimate of $0.14$. Since spatial-temporal coordinates were scaled into $[0, 1]$, the large temporal range indicates very strong temporal correlation, consistent with the covariance estimation in \cite{cao2023linear}. The tuning parameter $m$, controlling the conditioning set size in the Vecchia approximation of the log-likelihood of a partially censored GP \citep{cao2023linear}, was set to $50$, and the MC sample size was chosen as $10^4$ for the maximum likelihood estimation. Similar to \cite{cao2023linear}, chordal distance was used in our application study, as the chordal distance closely resembles the great-circle distance for locations within the United States. 

\begin{figure}[ht!]
\centering
\begin{subfigure}{.32\textwidth}
\centering
\includegraphics[width =.99\linewidth]{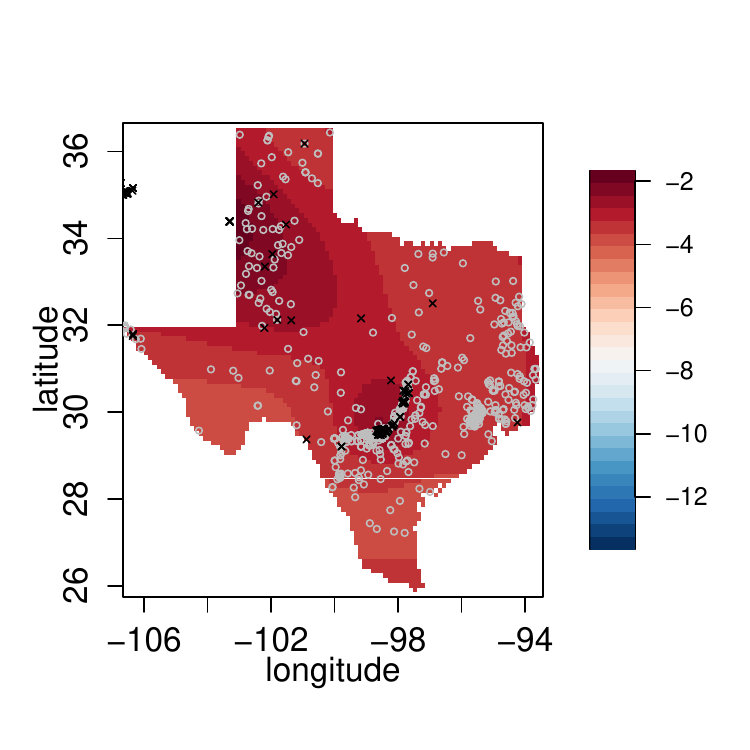}
\vspace{-2\baselineskip}
\subcaption{LOD-GP}
\end{subfigure}%
\hfill
\begin{subfigure}{.32\textwidth}
\centering
\includegraphics[width =.99\linewidth]{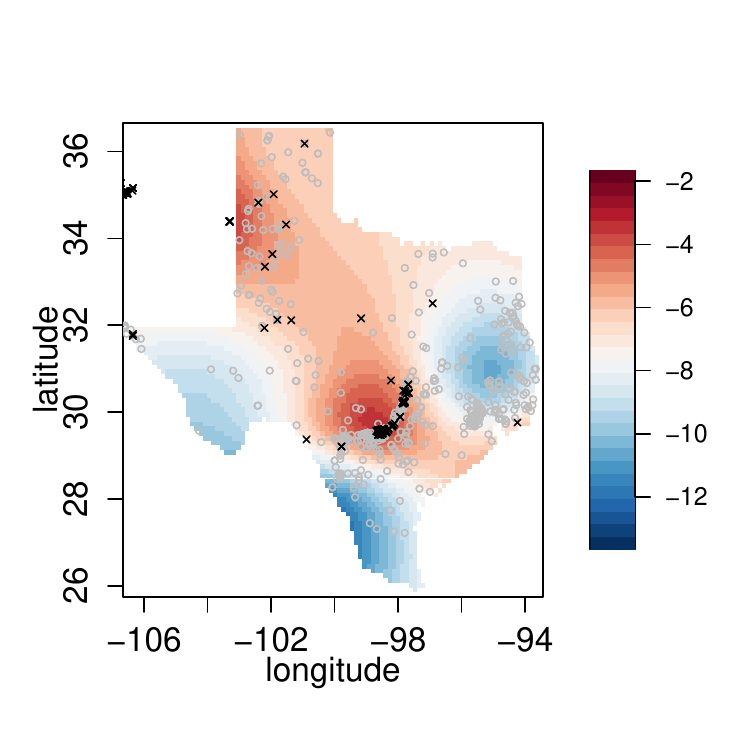}
\vspace{-2\baselineskip}
\subcaption{CSB}
\end{subfigure}%
\hfill
\begin{subfigure}{.32\textwidth}
\centering
\includegraphics[width =.99\linewidth]{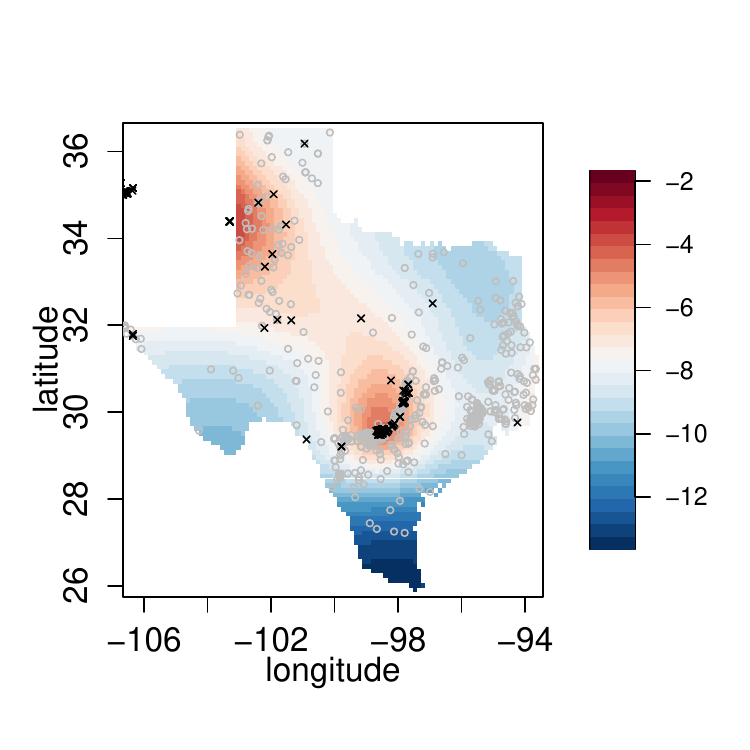}
\vspace{-2\baselineskip}
\subcaption{SNN}
\end{subfigure}%
\hspace*{\fill} \\

\hfill
\begin{subfigure}{.32\textwidth}
\centering
\includegraphics[width =.99\linewidth]{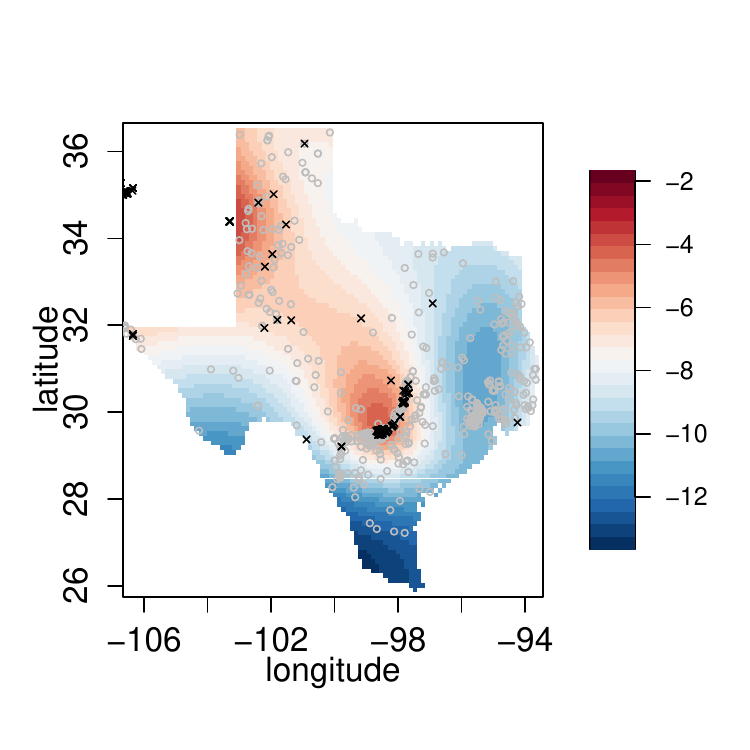}
\vspace{-2\baselineskip}
\subcaption{VMET}
\end{subfigure}
\hfill
\begin{subfigure}{.32\textwidth}
\centering
\includegraphics[width =.99\linewidth]{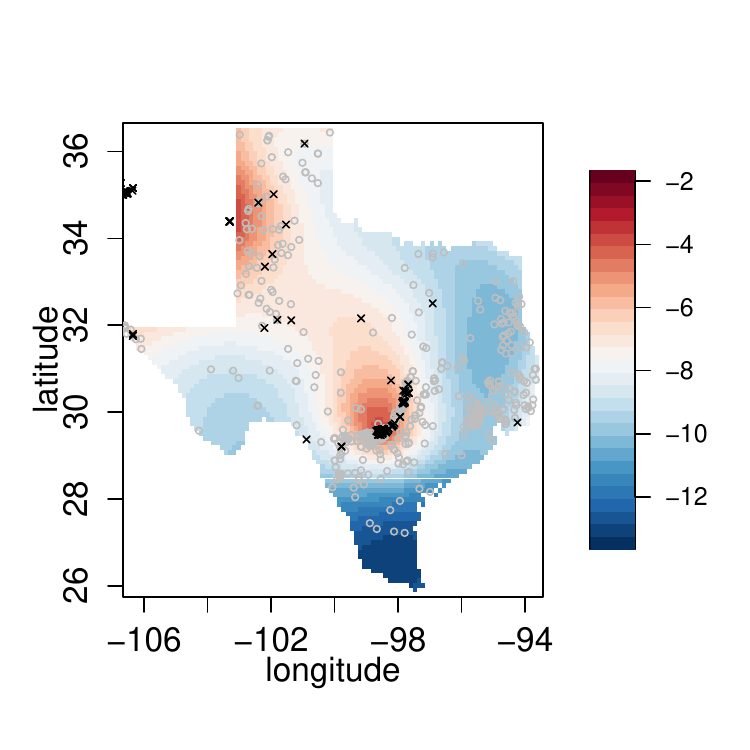}
\vspace{-2\baselineskip}
\subcaption{MET}
\end{subfigure}
\hspace*{\fill} \\
\caption{Heatmaps of posterior means of the latent GP over the Texas region generated by LOD-augmented GP, CSB, SNN, VMET, and MET}
\label{fig:TX_predict}
\end{figure}

First, we replicated the prediction over the Texas region as in \cite{cao2023linear} using LOD-augmented GP, CSB, SNN, VMET, and MET; the results are shown in Figure~\ref{fig:TX_predict}. 
There were $691$ censored responses located in the Texas region. LOD-augmented GP (LOD-GP) substituted the censored responses with LOD, while all other methods drew $N = 1{,}000$ samples over the $691$ spatio-temporal locations in Texas with censored responses conditional on all $n^{o} = 3{,}971$ observed responses in the U.S. For CSB, we used a burn-in period of $20{,}000$ iterations and thinning factor of $5$, same as in Section~\ref{sec:sim_study}. For SNN, we chose $m = 50$ but redefined the conditioning sets as the union of $25$ nearest neighbors from all censored responses and another $25$ nearest neighbors from all observed responses, because some locations were densely surrounded by censored responses, in which case including information from moderately distant observed responses may improve the accuracy of the conditional distribution. Based on $n^{o} = 3{,}971$ observed responses and $N = 1{,}000$ samples of the GP underlying the $691$ censored responses, ordinary kriging was used to make predictions over a $100 \times 100$ grid covering the Texas region. LOD-GP had significant positive bias, while CSB also had a certain degree of positive bias compared with SNN, VMET, and MET. Notice that MET could be used as the benchmark here if the GP is considered a reasonable assumption for the underlying spatial-temporal process. SNN, VMET, and MET all produced similar predictions.

\begin{figure}[ht!]
\centering
\hspace*{\fill} 
\begin{subfigure}{.45\textwidth}
\centering
\includegraphics[width =.99\linewidth]{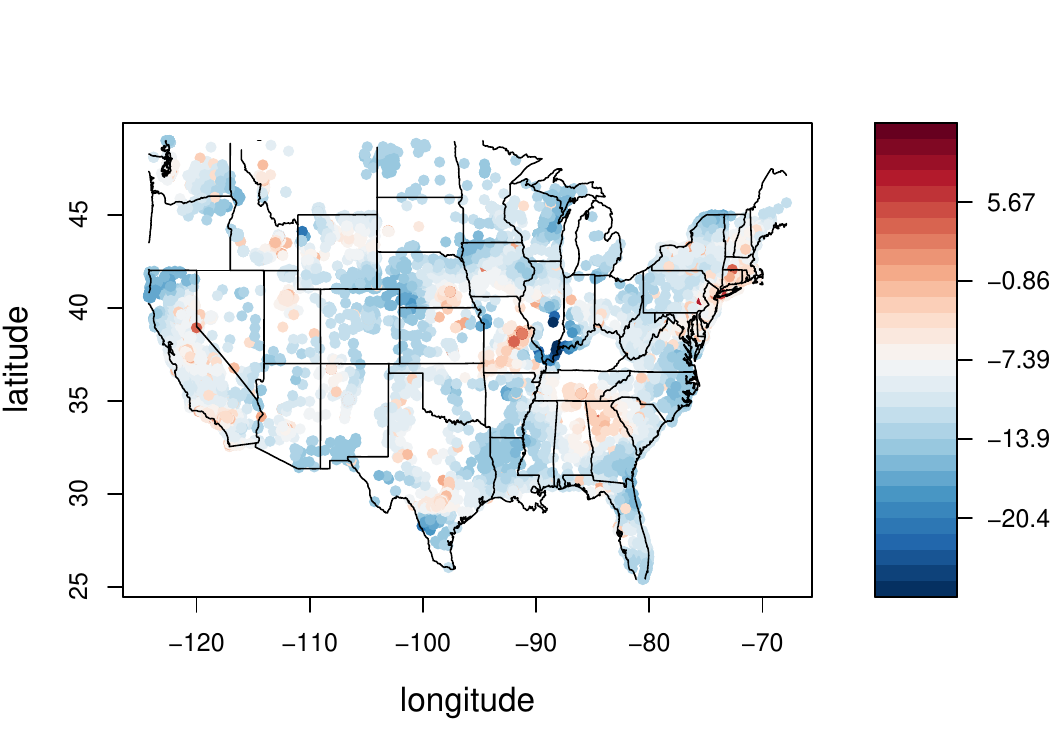}
\vspace{-1.5\baselineskip}
\subcaption{CSB}
\end{subfigure}
\hfill
\begin{subfigure}{.45\textwidth}
\centering
\includegraphics[width =.99\linewidth]{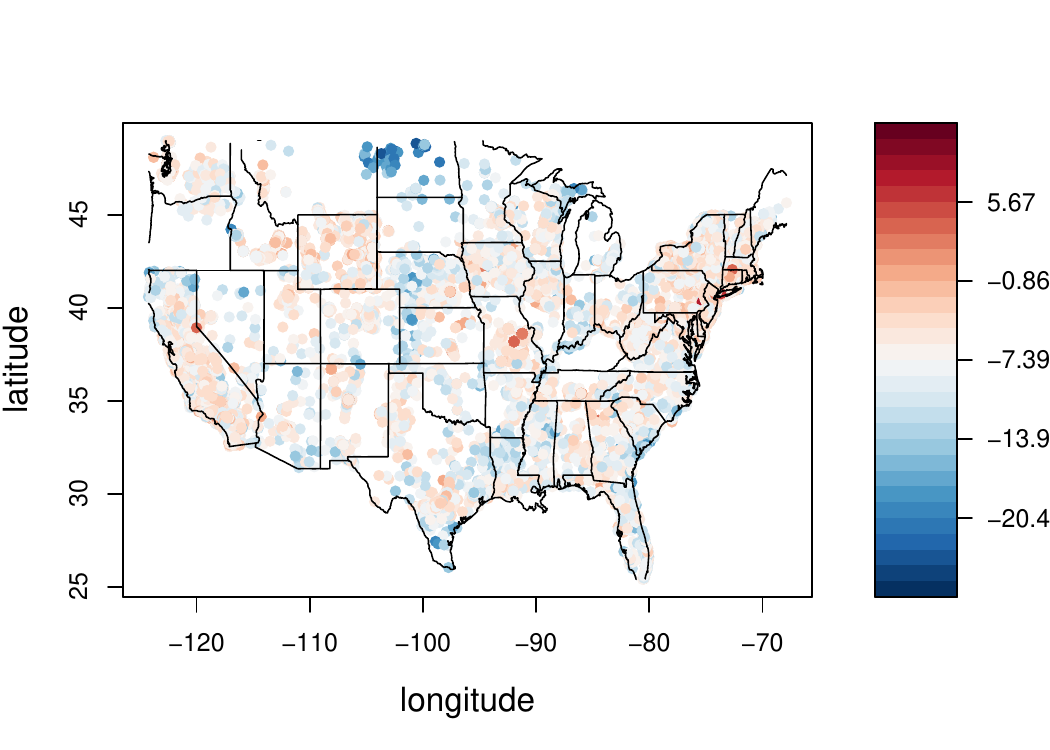}
\vspace{-1.5\baselineskip}
\subcaption{SNN}
\end{subfigure}
\hspace*{\fill} \\

\caption{Heatmaps of tetrachloroethylene concentrations across the United States predicted by CSB and SNN.}
\label{fig:US_predict}
\end{figure}

Next, we showcase the added capability of our proposed SNN over VMET and MET by generating joint samples over all $n^{c} = 20{,}730$ locations with censored responses. Figure~\ref{fig:US_predict} shows the heatmaps of the two samples generated by CSB and SNN, respectively.
Overall, CSB had a negative bias compared with SNN, especially visible in the central and eastern regions of U.S. Furthermore, the sample generated by CSB appeared rougher than that from SNN, which was also highlighted in the simulation study of Figure~\ref{fig:sim_study}. The roughness of CSB was likely caused by the conditional independence assumption between locations with censored responses given the responses over the latent (pseudo) locations \citep{sahoo2024computationally}.

\section{Conclusion}
\label{sec:conc}

We proposed a linear-complexity algorithm, coined SNN, for sampling high-dimensional TMVN distributions while maintaining a high fidelity. Different from other state-of-the-art methods \citep{botev2017normal, cao2023linear}, SNN is capable of maintaining a constant acceptance ratio in the accept-reject sampling regime. Hence, SNN can achieve linear complexity for each sample from the target TMVN distribution, instead of linear complexity for each sample from the proposal density as with \citet{cao2023linear}. Our proposed SNN is heuristically based on the screening effect \citep{Stein2011}, which has not been comprehensively tested for TMVN densities. However, both our simulation study and real-data application demonstrated the high fidelity of the samples generated by SNN, using those generated by \cite{botev2017normal}, the gold standard, as benchmark. \cite{sahoo2024computationally} is another recent method scalable to high-dimensional TMVN distributions by assuming conditional independence between censored responses, but its generated samples were shown to have lower fidelity than those generated by SNN. Furthermore, \cite{sahoo2024computationally} can be computationally challenging for GPs defined over $\mathbb{R}^{d}$ if $d \ge 3$ \citep{lindgren2023fmesher}.

There are two innovations underlying the SNN method. First, we use a sequential-nearest-neighbor approximation of the TMVN density function, which is reminiscent of the approaches in \cite{Nascimento2022} and \cite{cao2023linear} that apply the Vecchia approximation \citep{Vecchia1988} to the MVN density and probability functions, respectively. However, here we apply conditioning truncation to both equality-type conditions (i.e., \cite{cao2023linear}) and interval-type conditions (i.e., \cite{Nascimento2022}) and do not restrict the nearest neighbors being from previously ordered responses/locations. Second, we avoid sampling the analytically intractable marginal of a TMVN distribution through jointly sampling the TMVN distribution, which has been well established in the literature. For future research, we think one avenue is to explore if SNN can maintain high fidelity for general covariance structures beyond Mat\'ern kernels. Another possible avenue is to explore whether $\ba$ and $\bb$, in addition to $\{\bs_{i}\}_{i = 1}^{n}$, can be also used in constructing the neighbor sets $\{c(i)\}_{i = 1}^{n}$ to achieve even higher fidelity.

\section{Acknowledgments}

JC's research was supported by the Department of Mathematics at the University of Houston. 

\medskip \noindent
MK's research was partially supported by National Science Foundation (NSF) Grant DMS--1953005 and by the Office of the Vice Chancellor for Research and Graduate Education at the University of Wisconsin--Madison with funding from the Wisconsin Alumni Research Foundation.

\bigskip
\begin{center}
{\large\textbf{SUPPLEMENTAL MATERIALS}}
\end{center}

\begin{itemize}
    \item The R package implementing the SNN method can be found at: \url{https://github.com/JCatwood/nntmvn/tree/main}
    \item The code for replicating the results in this paper can be found at: \url{https://github.com/JCatwood/SNN-experiments/tree/main}
\end{itemize}

\bibliographystyle{apalike}
\bibliography{refs}

\end{document}